\documentstyle[aps,twocolumn,prl,epsf]{revtex}
\def\nn{\frac{1}{N_s}}
\def\w{\omega}
\def\e{\varepsilon}
\def\dps{\displaystyle}
\def\oin{\oint_{\cal C}}

\def\oizz{\oin\frac{dz}{2\pi i}e^{-\beta z}}

\def\intl{\int\limits}
\def\iwn{i \omega_n}

\def\intinf{\intl_{-\infty}^{\infty}}
\def\ul#1{\underline{#1}}
\def\kk{\ul{k}}

\begin{document}
\draft

 \twocolumn[\hsize\textwidth\columnwidth\hsize\csname
 @twocolumnfalse\endcsname
\title{Magneto-Resistance in the Two-Channel Anderson Lattice}

\author{F.~B.~Anders$^{(a)}$, Mark Jarrell$^{(b)}$, and D.L.
Cox$^{(a)}$}
\address{
$^{(a)}$Department of Physics,
The Ohio State University, Columbus, OH, 43210\\
$^{(b)}$ Department of Physics,
University of Cincinnati, Cincinnati, OH 45221
}

\date{September 20, 1996}

\maketitle

\widetext
\begin{abstract}
The paramagnetic phase of the two channel Anderson Lattice model in the
Kondo
limit is investigated in infinite spatial dimensions
using the non-crossing approximation. The   resistivity
exhibits a Kondo upturn with decreasing $T$, followed by a
slow decrease to a finite value at $T=0$. The decrease reflects lattice
coherence effects in concert with particle-hole symmetry breaking.  The
magneto-resistance obeys an approximate scaling
relation, decreasing towards coherent Fermi liquid behavior with
increasing field.
The magnetic field induces a Drude peak in
the optical conductivity.
\end{abstract}


]  

\narrowtext
\paragraph*{Introduction.}
Heavy fermion (HF) materials \cite{Grewe91} have been under
investigation
for two decades. Characteristic for these materials is a $100-1000$
fold enhanced electronic specific heat coefficient $\gamma(T)=C(T)/T$
and very large and
strongly temperature dependent resistivity $\rho(T)$ (of the order of
$100\mu\Omega cm$). In most of these inter-metallic compounds magnetic
or/and superconducting ground states are found. The physics of these
interesting anomalous metals is related to strongly
correlated electrons in $4f/5f$ orbitals.  Some of these systems
are well described as Landau Fermi liquids of massive quasi-particles.

Recently, a number of non-Fermi liquid (NFL) HF alloys have been found
which
display, e.g.,   logarithmically divergent
$\gamma(T)$\cite{Maple95}.
The superconducting HF compound
UBe$_{13}$ also has  NFL behavior in $\gamma$, and moreover
possesses a very large residual resistivity ($\approx 100 \mu\Omega
cm$) at
the superconducting transition  even in high quality samples
(as determined by a large $T_c$ and sharp resistive transition).
Such NFL behavior may be
possible from a two-channel Kondo model description of the
physics, such as has been proposed for UBe$_{13}$
\cite{Cox87} on symmetry grounds. In this picture,
{\it electrical quadrupole moments}
of the two-fold  non-magnetic $\Gamma_3$ ground state of the
$U$-ion
are screened by orbital motion of the
conduction electrons. Because the magnetic moment of the
electrons is a spectator to this process, there are two
screening channels.  Reversal of spin and orbital indices allows
for a two-channel magnetic Kondo effect for a
 $Ce^{3+}$ ion in a cubic
environment\cite{KimCox95}.
The two channel Kondo impurity model for these cases
\cite{Cox87,Cox93} has been investigated essentially
exactly with different techniques
\cite{Affleck91}. However,
little is known about the corresponding lattice model
\cite{Cox96,JarrellPanCoxLuk96}.

In this paper we present a solution of one particle properties of  the
two channel Anderson lattice model in infinite
spatial dimensions. To obtain the solution of the effective two
channel single impurity problem, the
Non-Crossing-Approximation (NCA)
\cite{Grewe83,Kuramoto83,Bickers87,Cox87}
is used. Although this method fails  to solve the infinite
dimension single channel Anderson lattice  at
temperatures less than the lattice Kondo-temperature $T^*$
\cite{KimKurKas87}, it works well in
the low temperature region of the two-channel model
\cite{Cox87,CoxRuck93,KimCox95}. The calculated  resistivity $\rho(T)$
agrees
well with recent
Quantum-Monte-Carlo (QMC) results \cite{JarrellPanCoxLuk96}.
For lower temperatures than accessible by QMC,
however, we find a decrease in $\rho(T)$ for decreasing $T$.
A strong negative
magneto-resistivity at  temperatures below the resistivity maximum in
an  applied magnetic field indicates a recovery of Fermi-liquid
behaviour. At the same time, an onset of a Drude peak in the optical
conductivity is found which is absent in the zero field solution.
We briefly discuss the possible relevance of these results to
experiment at the end.

\paragraph*{Theory.}
The two channel Anderson lattice Hamiltonian under investigation reads
\begin{eqnarray}
  \hat H  &= &\dps
\sum_{\alpha <i,j>} \frac{t^*}{\sqrt{d}}
c^\dagger_{i\alpha\sigma}c_{j\alpha\sigma}
+\sum_{i\sigma} E_\sigma X_{\sigma,\sigma}^{(i)}
\nonumber\\  &   &\dps
 +
\sum_{i\sigma\alpha} V\left\{
c^\dagger_{i\alpha\sigma}  X_{\alpha,\sigma}^{(i)}
+ h.c
 \right\} \;\; .
\end{eqnarray}
$X$ are the usual Hubbard operators,
$d$ being the spatial dimension, $i$ the lattice site, $t^*$ the
reduced hopping matrix
element of the conduction electron between nearest neighbours which
carry a spin $\sigma$ and a channel index
$\alpha=(1,2)$. The conduction electrons couple via hybridization
matrix element $V$ to the ionic many-body
states on each lattice site. The
symmetry  breaking magnetic field enters by a Zeeman term in $E_\sigma
= \e_f + g |\mu_b|\sigma H$. The  Zeeman splitting of the
conduction electrons only result in a shift of the band centers and
turn out to be a small correction. When $|E_\alpha- E_\sigma|$ is much
larger than the hybridization width $\Gamma_0 = \pi\rho V^2$, the
model can be mapped onto an two channel Kondo model \cite{Cox87} via
the Schrieffer-Wolff transformation \cite{SchriefferWol66}.

In the local approximation for the Anderson Lattice
\cite{KimKurKas87}, which is
equivalent to the limit $d\rightarrow\infty$ \cite{Georges96} with an
appropriate rescaling of the effective hopping \cite{MetznerVoll89},
we choose one $f$-site as an effective "'impurity"' site which is
self-consistently embedded in an effective medium reflection the rest
of lattice. While in a single impurity problem only the bare medium
$\Delta_{\alpha\sigma}^{0}(z) =
N_s^{-1}\sum_{\kk}d_{\alpha\sigma}(\kk,z)
=N_s^{-1} \sum_{\kk}{V^2}(z-\e_{\kk \alpha\sigma})^{-1}$
enters, in the lattice the condition that the local $f$-Green's
function (GF) has to be
equal to the $\kk$-summed lattice GF
\begin{equation}
\label{eqn-scc-doo}
 \nn \sum_{\kk}
\frac{1}{1- \tilde F_{\alpha\sigma}(z)\left(d_{\alpha\sigma}(\kk,z) -
\tilde\Delta_{\alpha \sigma}(z)\right)}
= 1 \;\; .
\end{equation}
where $\tilde\Delta_{\alpha\sigma}(z)$ is the self-consistent one body
self energy of the local "impurity" propagator
$\tilde F_{\alpha\sigma}(z)
=\ll\hat X_{\alpha,\sigma}^{(i)}|\hat X_{\sigma,\alpha}^{(i)}\gg (z)
$.
The effective hybridization width
$\tilde\Gamma_{\alpha\sigma}(\w) = \frac{\Im
m}{\pi}\tilde\Delta_{\alpha\sigma}(\w-i\delta)$ enters then the
local two channel impurity problem.  The self-energy of the conduction
electrons is given by
\begin{equation}
\Sigma_{\alpha\sigma}(z)= \frac{V^2 \tilde
F_{\alpha\sigma}(z)}{1 + \tilde
F_{\alpha\sigma}(z)\tilde\Delta_{\alpha\sigma}(z)}
\;\; .
\label{equ-c-self}
\end{equation}

In the single channel case, the unitarity limit of the $T$-matrix
$\tilde T(z)=V^2\tilde F(z)$ leads to the Fermi-liquid behaviour of the
conduction
band self-energy $\Sigma_{\sigma}(z) \propto a T^2 + b w^2$.
Since the value of the $T$-matrix  at the chemical
potential and $T\rightarrow 0$ is smaller than the unitarity limit
 in the two(multi)-channel case
Eqn.(\ref{equ-c-self}) tells us immediately, that
the corresponding conduction band self-energy for the exact solution in
the
paramagnetic phase of the lattice has to be finite. This has been
recently called an {\em incoherent metal}
\cite{Cox96,JarrellPanCoxLuk96}. The physical origin is the following:
the local spin is over-compensated by  two conduction electron
spins. On each lattice site a residual free thermodynamically
fluctuating degree of freedom (DOF) acts as a scatterer for conduction
electrons. An
residual entropy of $1/2\log(2)$ per site is associated with this DOF
freedom which has been interpreted as a free Majorana fermion
\cite{EmeryKiv92}.  The  finite self-energy
yields a finite value for $\rho(T\rightarrow 0,H=0)$. Since in
translational invariant system a vanishing dc
resistivity is expected this indicates that the paramagnetic state is
{\em not} the ground state of the two-channel Anderson lattice.

In the absence of a magnetic field the NCA equations of the effective
impurity
\begin{eqnarray}
\Sigma_\alpha(z) & = & \frac{1}{\pi} \sum_\sigma \intinf
\tilde\Gamma_{\alpha\sigma}(\e) f(\e)\tilde P_\sigma(z+\e)
\label{eqn-sig-a}
\\
\Sigma_\sigma(z) & = & \frac{1}{\pi} \sum_\alpha \intinf
\tilde\Gamma_{\alpha\sigma}(\e) f(-\e)\tilde P_\alpha (z-\e)
\label{eqn-sig-s}
\end{eqnarray}
are equivalent to a resonant level system with an effective Anderson
width $\tilde\Gamma_0= 2\Gamma_0$. The NCA pathology
 in the local GF
becomes the physical Abrikosov-Suhl-resonance (ASR) in the two channel
case \cite{Cox87,KimCox95,Cox93}. The NCA threshold exponents of the
effective ionic propagators $\tilde P$ \cite{MuellerHartman84} have
the exact value of $1/2$ calculated within a conformal field theory
approach\cite{AffleckLud93}. In limit of infinite spin  $N$ and channel
$M$ degeneracy with a fix ratio $N/M$ the NCA becomes exact
\cite{CoxRuck93}.
The effective local  GF is given by the convolution
\begin{equation}
\tilde F_{\alpha\sigma}(\iwn) = \frac{1}{\tilde Z_f}\oizz \tilde
P_\alpha(z)\tilde P_\sigma(z+\iwn) ,
\end{equation}
where $\tilde Z_f$ is the effective local partition function.
Even though higher order vertex corrections \cite{Anders95}
will modify the spectral distribution, the leading physical effect and
the correct thermodynamics is captured correctly within the
Eqns.(\ref{eqn-sig-a}) and (\ref{eqn-sig-s}).
Since the saturation value of the effective site $T$-matrix is half
the unitary limit  no pseudo-gap develops in the quasi-particle
spectrum as in the onc-channel lattice
\cite{GrewePruKei88}.

In infinite spatial dimensions the vertex corrections in the
two-particle propagators vanish \cite{MuellerHartman89} and the
conductivity itself is a
$1/d$ correction which can be calculated by evaluating the lowest
order bubble diagram \cite{PruschkeCoxJar93}, given by
\begin{eqnarray}
\sigma_{\alpha}(\w) & = & \dps  A
\intinf d\w'\frac{[f(w')-f(w+w')]}{\w}
\intinf d\e \rho_0(\e)
\nonumber \\
&& \sum_{\sigma} \dps
\Im m
G^{(c)}_{\alpha\sigma}(\w'-i\delta,\e)\Im m G^{(c)}_{\alpha\sigma}(\w
'+ \w-i\delta,\e)  \;\; ,
\label{equ-opti-cond}
\end{eqnarray}
which can be written as an integral over four complex error function;
$A = \pi e^2 a^2 t^{*^2}N (h d Vol)^{-1} =t^* \omega_p^2/(4\pi)$, the Gaussian density of
states $\rho_0(\e)$ \cite{MetznerVoll89},
$G^{(c)}_{\alpha\sigma}(z)$  the conduction electron GF, and $a$ the
lattice
constant of the $d$-dimensional hypercube.
The $f$-electrons do not contribute to the conductivity since the
hybridization is $\kk$ independent. The dc-conductivity is obtained by
the limit $\sigma_{dc}(T) = \lim\limits_{\w\rightarrow
0}\sigma(\w,T)$.

\paragraph*{Results.}
To obtain a self-consistent solution of the
lattice problem, {\em (i)}  the effective hybridization with
$\tilde\Gamma(\e)$ has
been treated with the same accuracy as the threshold singularity of
the ionic propagators in Eqns.(\ref{eqn-sig-a}) and
(\ref{eqn-sig-s}), and {\em (ii)} only 10\% of the calculated change of
$\tilde\Gamma(\e)$ is added in each lattice iteration step, i.e.~the
lattice is switched on adiabatically. The error in norm of the $\tilde
P(z)$ reaches $0.01\%$, the sum rule for the self-energy is obeyed
within 0.02\%
and the maximum norm
$max\{|\tilde\Gamma_{n}(\e)-\tilde\Gamma_{n-1}(\e)|\} <
10^{-8}$. All energies, if not otherwise stated, are measured in the
original Anderson-width $\Gamma_0$. We chose $\e_f= E_\sigma-E_\alpha
= -3\Gamma_0$
in the absence of $H$ and $t^*=10\Gamma_0$ with a band center at
$\w=0$. $T_K= 0.016\Gamma_0$ \cite{tkondo}.

\begin{figure}[t]

\epsffile{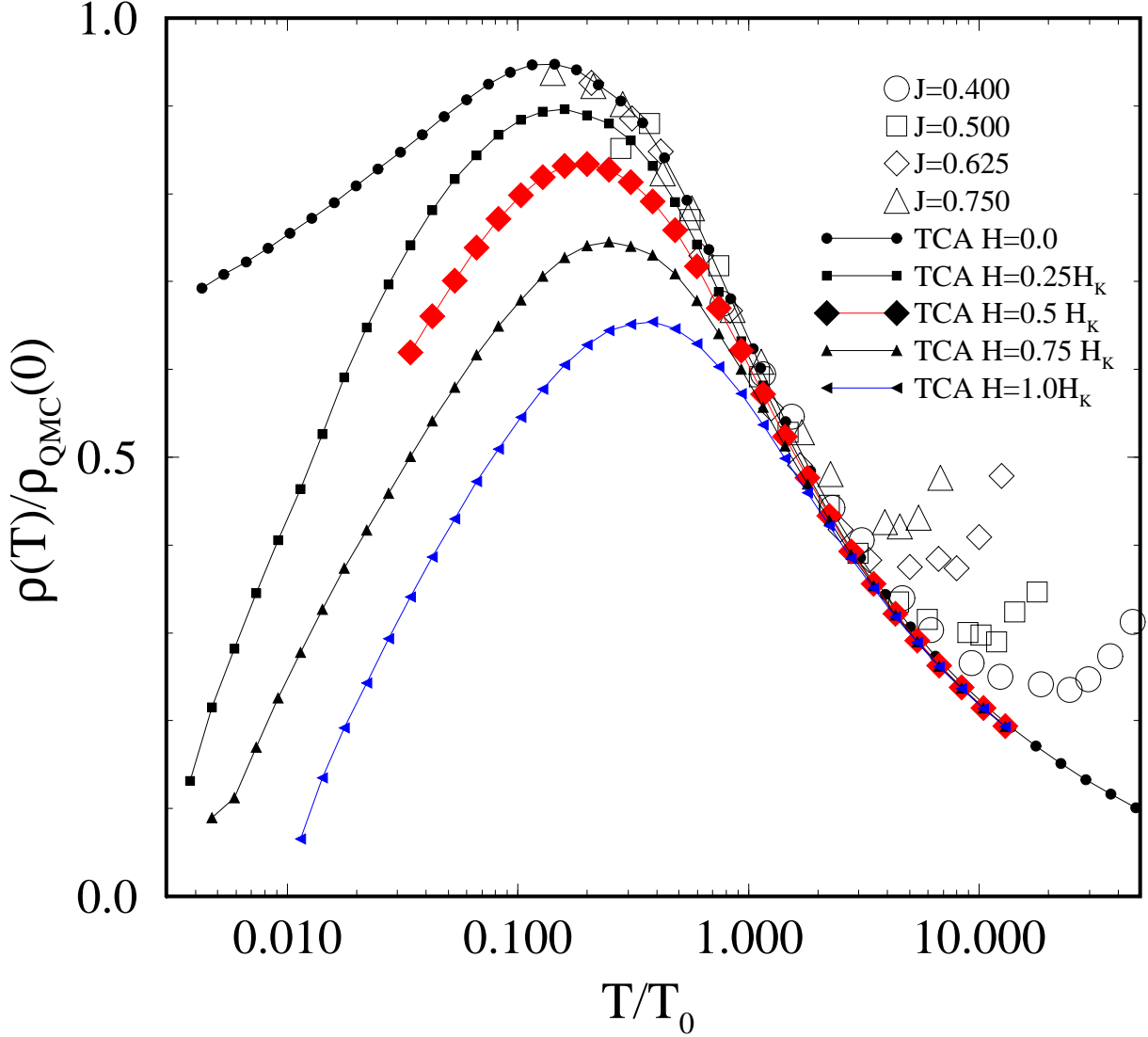}
\caption{Resistivity for the two channel Anderson lattice model (TCA)
vs.~temperature for different magnetic fields. We have  normalized to
the
estimated $\rho(T=0)$ values of the Quantum-Monte-Carlo (QMC) data,
temperature for the same parameters. The open symbols are the
QMC results for different $J$ in the two-channel Kondo
lattice model. The inset shows the scaling of the position of the dc
maxima vs $H$.
 Parameters: $\e_f =-3\Gamma_0, t^*=10\Gamma_0, T_0=0.02\Gamma_0$.
}
\label{fig-1}
\end{figure}

In Fig.~\ref{fig-1} $\rho(T,H))$ normalized to the estimated
$T\rightarrow 0$ value of the QMC data \cite{JarrellPanCoxLuk96} is
shown  for different values of the an applied magnetic field measured
in units of $H_K = k_BT_K/(g\mu_B)$. We have fix the lattice scale
$T_0 = 1.3T_K$ \cite{tkondo} by matching the QMC resistivity data.
The agreement with the higher
temperature  data for the Kondo lattice - the open symbols - is
excellent. Nevertheless, the resistivity has a maximum and slowly
decreases with decreasing temperature, as expected from a lattice
calculation. The calculations are done slightly below half-filling and
with fixed chemical potential. At
half-filling in the Kondo-regime an the analytical solution obtained
with an artificial Lorentzian density of states \cite{Cox96} predicts
a resistivity of $\rho(T) = \rho(0)(1-a\sqrt{T})$ at low
temperature.  While for
$H=0$ still a positive intercept at $T=0$ is expected consistent with an
infinitly degenerated ground state,
clearly a crossover to Fermi-liquid is found within an applied
field. The NCA pathology \cite{MuellerHartman89}
prevents  access to the Fermi-liquid  regime when
$T\rightarrow 0$ \cite{KimKurKas87}.
Evaluation the constants for $A$ in
Eqn.~(\ref{equ-opti-cond}), assuming a lattice constant of 5\AA\ and 2
electrons per unit cell in a three dimensional lattice gives a
resistivity prefactor of $\approx 12.6 \mu\Omega cm/\Gamma_0^2$ which
leads to a resistivity maximum of $\approx 250\mu\Omega cm$ using our
absolute maximum of $20\Gamma_0^2$. This is  very close to the
experimentally found value of $\approx 190 \mu\Omega cm$ for UBe$_{13}$
\cite{WillisThoSmiFis87}.

\begin{figure}[t]

\epsffile{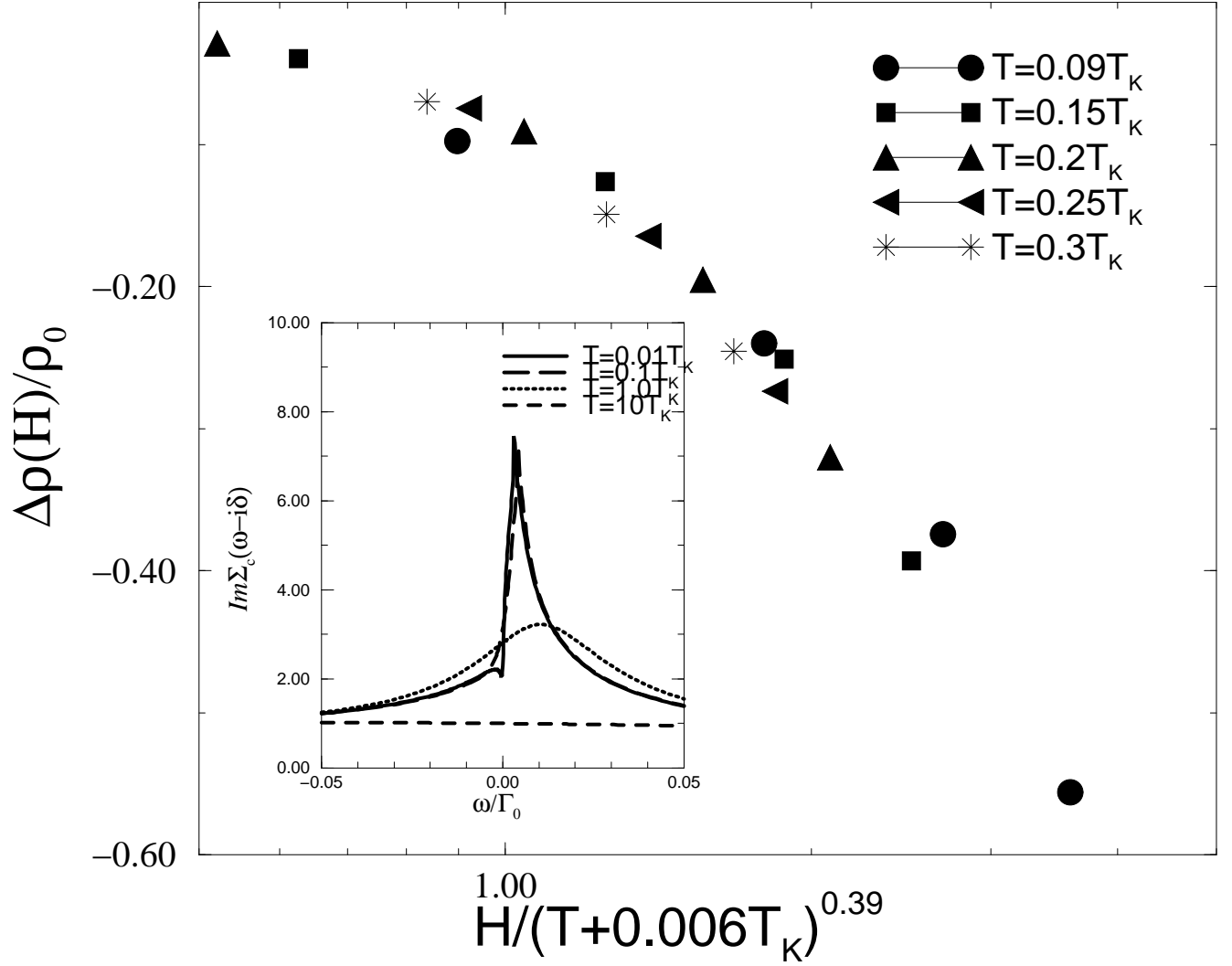}
\caption{Magneto-resistance
$(\rho(T,H)-\rho(T,0))/ \rho(T,0)$ vs scaling variable $x=H/(T+0.006
T_K)^{0.39}$. {\em Inset:}
Imaginary part of the conduction band self-energy vs. frequency in the
vicinity of the chemical potential $\mu=0$ for four different
temperatures. Parameters: $\e_f =-3\Gamma_0, t^*=10\Gamma_0$.
}
\label{fig-2}
\end{figure}

Motivated by the experimental data for $\rho(T,H)$ for UBe$_{13}$
\cite{AndrakaSte94}, we have attempted to scale our $\rho(T,H)$ data
with the {\em Ansatz} $\Delta\rho/ \rho =
[\rho(T,H)-\rho(T,0)]/ \rho(T,0) \propto f(H/(T+ T^*)^{\beta})$.  While
for the impurity model, we expect $T^*
=0., \beta = 1/2$, we find approximate scaling for $T^*=0.006T_K,
\beta=0.39$, as ploted in  Fig.~\ref{fig-2}.
The inset of the figure shows the
imaginary part of the conduction electron self-energy
$\Sigma_c(\w-i\delta)$ (\ref{eqn-sig-s}) for $H=0$ is plotted for four
different temperatures in Fig.~\ref{fig-2}. It shows a shift of the
maxima away from the
chemical potential in this metallic regime.  Very close to $\w=0$ a
very small onset of coherence is observed for $T\rightarrow 0$, but
the relaxation rate remains of the order of $2\Gamma_0$.

\begin{figure}[t]
\epsffile{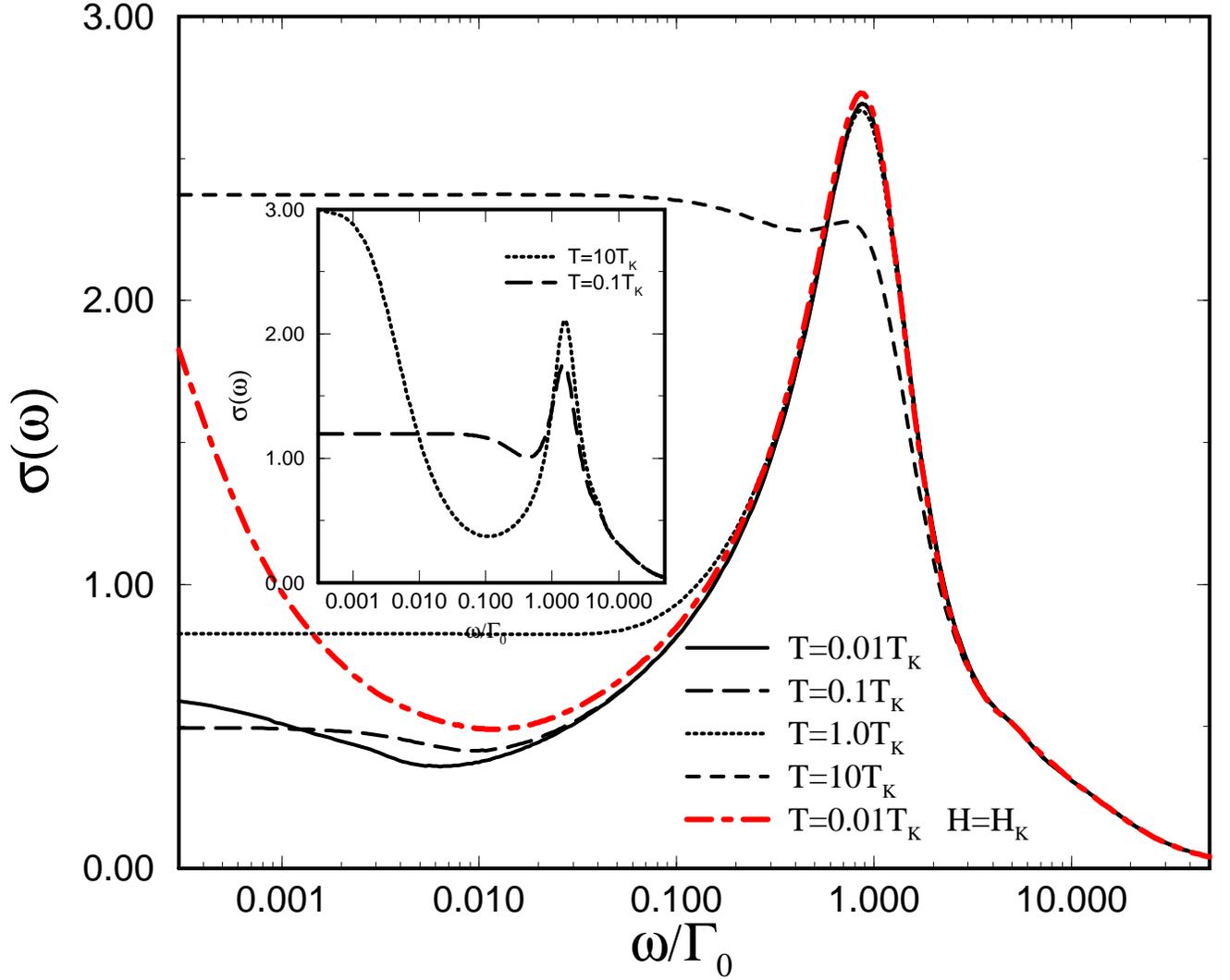}
\caption{
Optical conductivity per channel in units of $\omega^2_p/(4\pi)$
vs.~frequency for covering three decades in temperature.
In the inset the  corresponding curves for the single channel PAM
for the two highest temperatures are show in the same units.
The dash-dotted line is calculated with $H=H_K$.
 Parameters: $\e_f =-3\Gamma_0, t^*=10\Gamma_0$.
}
\label{fig-opti}
\end{figure}

Now we  focus on the optical conductivity, displayed in figure
\ref{fig-opti}. The large peak at $\approx 0.9\Gamma_0$ results from
high energy charge excitations.
With decreasing temperatures the optical conductivity
develops a pseudo-gap. The $f$-sum rule relates the integrated
optical conductivity
\begin{equation}
\int\limits_0^\infty \sigma_{\alpha}(\w) d\w
=
\frac{\pi^2e^2 a^2 t^{*^2}}{h}\frac{1}{Vol}
\sum\limits_\sigma \ll \hat T_x\gg
\;\; \;\propto \frac{1}{d}
\end{equation}
to the average kinetic energy in the direction of the current flow
\cite{Maldague77},
which is checked numerically. It
indicates a shift of small amount of spectral weight to higher
frequencies. This can be seen clearly in the figure by comparing the
$T=10T_K$ and the
$T=T_K$  curve (note that the logarithmic plot overemphazies the area
of the gap).
At low temperatures a small increase in $\sigma(\w)$ can be observed
when $\w\rightarrow 0$. Nevertheless, no clear Drude
peak is seen even for $T=0.01T_K$, one decade lower than the observed
maximum in $1/\sigma(0,T)$.
However, in a magnetic field of $H=H_K$ a low frequency "Drude"-peak
develops
again, consistent with the return to Fermi-liquid behaviour suggested
in $\rho(T,H)$. Note that in the single channel Anderson lattice a
clear Drude peak develops already a little below the maximum in
$1/\sigma(0,T)$.

\begin{figure}[t]
\begin{center}
\epsfysize 100mm
\epsffile{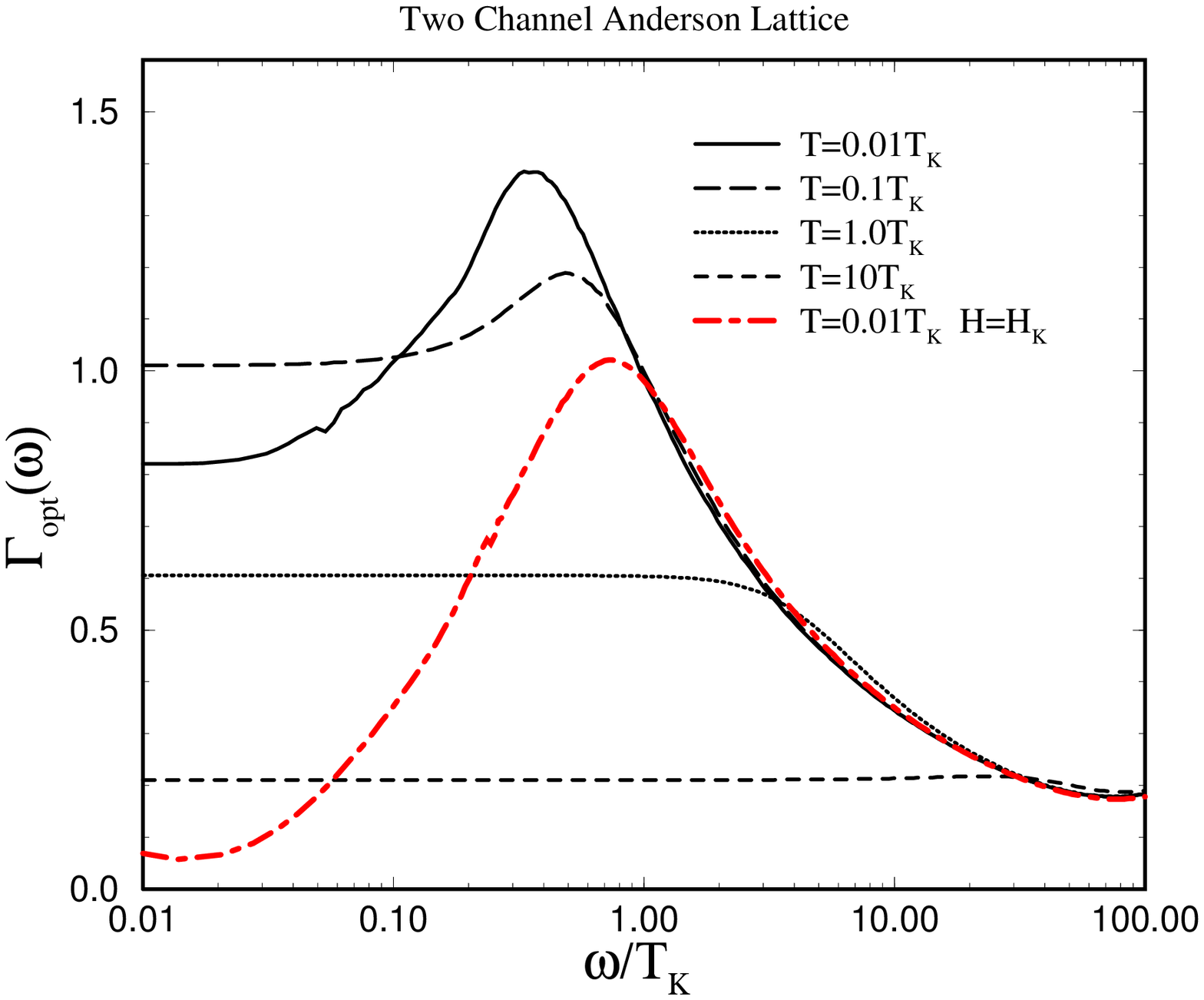}
{\hfill (a)}

\epsfysize 100mm
\epsffile{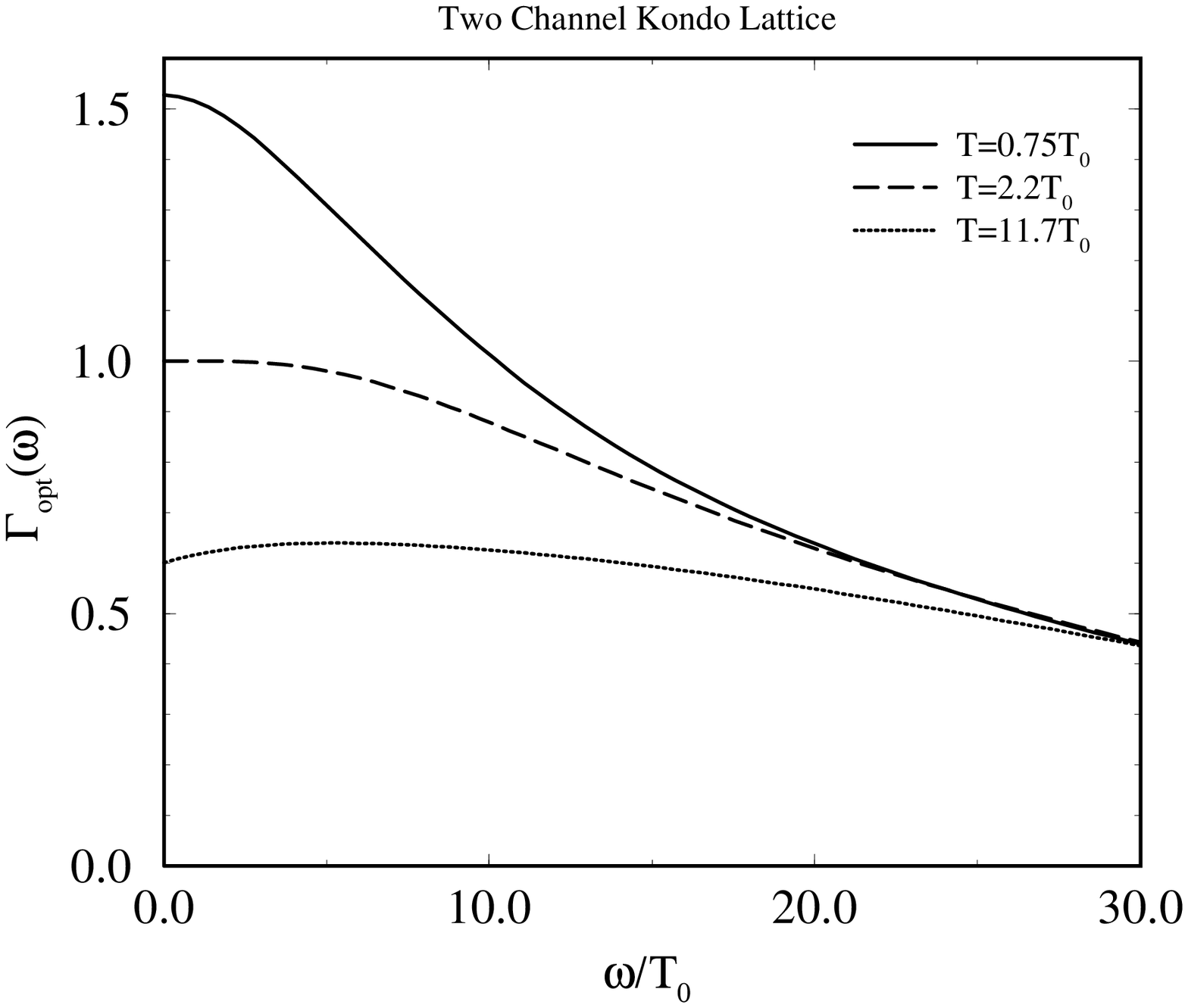}
{\hfill (b)}
\end{center}

\caption{
Relaxation~rate~$\Gamma_{opt}(\w)$ in units of $\Gamma_0$ (a) per
channel for the two channel 
Anderson lattice measured   versus $\omega/T_K$
 (Parameters: $\e_f =-3\Gamma_0, t^*=10\Gamma_0$)
and (b) QMC data for the two channel Kondo lattice (Parameter:
$J= 4\Gamma_0$, $T_0 = 0.281\Gamma_0$).
}
\label{fig-gamma}
\end{figure}

Using the Kramers-Kronig relation the imaginary part of the optical
conductivity has been calculated. With the phenomenological {\em
Ansatz}
\begin{equation}
\sigma_{opt}(\w) = \sigma(\w) + i\sigma''(\w) =
\frac{\w_p^2}{4\pi} \frac{1}{\Gamma_{opt}(\w) -i\w (1+\lambda(\w))}
\end{equation}
 the dynamical optical relaxation rate
$\Gamma_{opt}(\w)$ and the dimensionless mass enhancement factor
$\lambda(\w)$ have been  determined. In Fig.~\ref{fig-gamma}(a)
$\Gamma_{opt}(\w)$ is plotted for the same parameters as in
Fig.~\ref{fig-opti}. While for temperatures $T>0.1T_K$
$\Gamma_{opt}(\w)$ is nearly frequency independent for low
frequencies, at  the lowest temperature $\Gamma_{opt}(0)$ has
decreased reflecting the decrease of the dc-resistivity. The low
frequency behaviour has an exponent slightly lower then $n=1$.
We emphasize the {\em difference} between
$\Im m\Sigma_c(\w)$  and $\Gamma_{opt}(w)$: the first is a true {\em
one-particle} relaxation rate, the second, however, reflects the
two-particle nature of the energy absorption process associated with
electrical charge transport. Generally, only for a Fermi-liquid at
very low temperatures and frequencies should $\Gamma_{opt}=-2\Im
m\Sigma_c(\w)$.
 In an applied  magnetic field  $H=H_K$ and low temperatures
 $T=0.01T_K$
the optical relaxation rate $\Gamma_{opt}(0)$ shows the expected trend
to Fermi-liquid behaviour.

Fig.~\ref{fig-gamma}(b) shows the quantum Monte Carlo (QMC) calculation
of
$\Gamma_{opt}(\w)$ for the two-channel Kondo lattice at particle hole
symmetry.  A strict quantitative
calculation is not possible because: (i) of the overlap of $T_0$ and
the Kondo
interaction $J$ to within
an order of magnitude in the QMC calculations (in the Anderson model
calculations of this paper $T_0$ is well separated from high energy
scales), and (ii) because the particle-hole symmetry removes the
non-monotonicity
experienced in the NCA calculations.  Modulo these concerns, the
separate
calculations agree qualitatively in the overlapping temperature and
frequency regions.

{\em Comparison to Experiment:}
As mentioned earlier, $\rho(T,H=0)$ is reminiscent in form and
magnitude to UBe$_{13}$ \cite{WillisThoSmiFis87,AndrakaSte94}.
The magneto-resistance also resembles that of UBe$_{13}$, though our
scaling form in detail is different. However, a strict comparison is
not possible, since assuming a quadrupolar Kondo model applies to
UBe$_{13}$, we should rather split $|\alpha i>$ states (order H) and
quadratically split $|\sigma i>$ stated (van Vleck
processes). Additionally, recent experiments suggest a possible
 U$^{3+}$-U$^{4+}$ configuration degeneracy with is lifted
with Th substitution \cite{Aliev95}. We refer the necessary
intermediate valance
calculation to a future work. In this case, a crossover from NFL to
Fermi-liquid physics is still expected. Our
$\sigma(T,\w),\Gamma_{opt}(\w)$
calculations are very compatible with data for the alloys
Y$_{0.8}$U$_{0.2}$Pd$_{3}$ and Th$_{1-x}$U$_{x}$Pd$_{2}$Al$_{3}$
\cite{Degiorgi95},  as well as the compound UBe$_{13}$\cite{Bonn87}.
Because of the incoherent normal metal phase, we
expect little qualitative difference between these more dilute alloys
and the lattice.  For UBe$_{13}$, the existing optical data only go to
50 cm$^{-1}\simeq 5-6k_BT_K$ in frequency\cite{Bonn87}, and it is
clearly
desireable to extend these measurements to lower frequencies.
We remark that  for
Th$_{1-x}$U$_{x}$Pd$_{2}$Al$_{3}$, if a hexagonal quadrupolar
Kondo picture applies, a $c$-axis magnetic field will split the
$|\sigma i>$ levels \cite{Cox93}, permitting comparison to our
calculations.  In addition, it should be interesting to test whether a
magnetic
two-channel lattice picture applies to CeCu$_2$Si$_2$\cite{KimCox95}
 and thus have detailed
optical conductivity measurements carried out in applied field for this
system.

 We would
also like to thank P.~Coleman, A.~Millis and M.B.~Maple for organizing
a
stimulating workshop on {\em NFL behaviour in solid} at the
Institute for Theoretical Physics, where part of the work was
performed. Especially, we thank P.
Coleman for suggesting a careful analysis of the magnetotransport.
One of us (FBA) also like to acknowledge encouraging
comments by H.~Castella, D.~Vollhardt and
N.~Grewe. This~work~was~supported~by the Deutsche
Forschungsgemeinschaft  in part by the National Science Foundation
under Grant No. PHY94-07194, and the US Department of Energy, Office of
Basic Energy Science, Division of Materials Research (FBA and DLC), and
by NSF grants DMR-9406678 and DMR-9357199 (MJ).   Quantum Monte Carlo
calculations were carried out with a grant of supercomputer time from
the Ohio Supercomputer Center.
The NCA calculations were performed partially on a Pentium
driven LINUX laptop.


\onecolumn
\clearpage

\begin{references}
\bibitem{tkondo}
Since there is not an unambiques analytical formula for the  lattice
Kondo temperature $T^*$ we used $T_K=0.016\Gamma_0$ as a reference
temperature for our NCA calculations.
\end{references}
\end{document}